\documentclass[useAMS,usenatbib]{mn2e}
\usepackage{epsfig,amssymb,color}
\newcommand{\red}{\protect\color{red}}
\newcommand{\blue}{\protect\color{blue}}
\newcommand{\black}{\protect\color{black}}

\renewcommand\blue\red
\renewcommand\red\black

\title[Density and metallicity of the circumgalactic gas]{
Density and metallicity of the Milky-Way circumgalactic gas}
\author[S.~Troitsky]{%
Sergey Troitsky
\thanks{E-mail: st@ms2.inr.ac.ru}\\
Institute for Nuclear Research of the Russian Academy of
Sciences,
60th October Anniversary prospect 7a, 117312, Moscow, Russia,\\
and\\
Moscow Institute for Physics and Technology,
Institutskii per.\ 9, 141700, Dolgoprudny, Moscow Region, Russia
}
\begin{document}
\date{\blue 2017 February 02; \black in original form 2016 July 19}
\pagerange{\pageref{firstpage}--\pageref{lastpage}} \pubyear{2016}
\maketitle
\label{firstpage}
\begin{abstract}
The halo of the Milky-Way circumgalactic gas extends up to the virial
radius of the Galaxy, $\sim 250$~kpc. The halo properties may be
deduced from X-ray spectroscopic observations and from studies of the
ram-pressure stripping of satellite dwarf galaxies. The former method is
more precise but its results depend crucially on the assumed metallicity
of the circumgalactic gas; the latter one does not need these
assumptions. Here, the information from both approaches is combined to
constrain observationally the gas metallicity and density as functions of
the galactocentric distance. It is demonstrated that the \blue
two kinds of data could be reconciled if the \black metallicity
decrease\blue{}d \black \red to $Z\sim 0.1Z_{\odot}$ \black in the outer
parts of the extended halo. \red The corresponding gas density profile is
rather flat, falling as $r^{-(0.45 \dots 0.75)}$ at large galactocentric
distances $r$. \black
\end{abstract}


\section{Introduction}
\label{sec:intro}
Recently, considerable attention has been attracted to studies of gas
coronae of galaxies, that is of reservoirs of gas extending up to the
galaxies' virial radii. This circumgalactic gas represents, thanks to the
large volume it fills, a substantial contribution to the mass budget of a
galaxy. This gaseous corona, or extended halo, of the Milky Way has
attracted particular interest because of the ``missing-baryon'' problem,
see e.g.\ \citet{DM}, the apparent lack of baryons in our Galaxy compared
to the amount expected, on average, from cosmology. On the other hand,
interactions of cosmic rays with this circumgalactic gas have been
considered as a possible source of important contributions to the diffuse
gamma-ray \citep{Hooper} and neutrino \citep{Aha} backgrounds.

In the Milky Way, this reservoir of gas reveals itself in observations in
two ways. First, the ram pressure of the gas strips dwarf satellite
galaxies, whose orbits lay within the corona, from their own gas
\citep{dwarfs250kpc}. Second, the presence of the hot gas may be seen in
X-ray spectra, either as zero-redshift absorption lines for extragalactic
sources, or as emission lines in the blank-sky spectrum, see e.g.\
\citet{gas1,MB2013,1412.3116}. Taken at face
value, the gas density profiles derived by these two methods are
inconsistent with each other. However, the spectroscopic approach is based
on observations of spectral lines of oxygen, which is only a tracer of the
full amount of gas. As a result, the gas density obtained by the
spectroscopic method is very sensitive to unknown chemical composition of
the gas, usually encoded in its metallicity $Z$. The aim of this work is
to depart from simplified ad hoc assumptions about the metallicity of the
Galactic corona and to use spectroscopic and ram-pressure results jointly,
which allows us to constrain values and profiles of density and
metallicity of the circumgalactic gas simultaneously, so that the
agreement between all data is maintained.

The rest of the paper is organized as follows. In Sec.~\ref{sec:obs},
observational constraints on the density of circumgalactic gas are
discussed in detail. In particular, in Sec.~\ref{sec:X-ray}, X-ray
spectroscopic results are discussed and their dependence on the
assumptions about metallicity is recalled. Sec.~\ref{sec:ram} discusses
constraints from ram-pressure stripping of the Milky-Way satellites;
a combined fit of the most precise of these bounds is presented.
Other constraints are briefly mentioned in Sec.~\ref{sec:DM}.
Section~\ref{sec:comb} contains the main results of the paper and presents
a combination of the constraints, allowing to determine both the density
and the metallicity of circumgalactic gas in a joint fit by means of
statistical marginalization. These results are discussed and compared to
previous works in Sec.~\ref{sec:concl}.

\section{Observational constraints}
\label{sec:obs}

\subsection{X-ray spectroscopy}
\label{sec:X-ray}
Observations of distant extragalactic sources in X rays reveal O{\small
VII} and O{\small VIII} absorption lines which, unlike others, are
positioned at the redshift $z\simeq 0$, indicating that they originate
from hot absorbing gas near the observer. Similar, but emission, lines
were found in spectra of the sky obtained from directions where no sources
are present. A large amount of observations were interpreted as an
evidence for an extended circumgalactic gas halo. Having high statistical
significance, these results suffer however from considerable systematic
uncertainties related to the fact that the observed oxygen is only a
tracer of the full amount of gas, expected to be mostly hydrogen.

Suppose that the concentrations of heavy chemical elements in the gas
follow those in the Sun, so that they are encoded in a single parameter,
metallicity.
The true total electron density $n_{e}$ of the gas at the
galactocentric distance $r$ is related to the value $n_{e}'$ obtained in
this approach (see \citet{MB2013} for a detailed discussion) as
\begin{equation}
n_e (r) = n_{e}' (r) \left(\frac{Z(r)f(r)}{Z'
f'} \right)^{-1},
\label{Eq:*}
\end{equation}
where $Z$ is the metallicity and $f$ is the ionization
fraction, while $Z'$ and $f'$ are their assumed values. Previous works used
$Z'=0.3 Z_{\odot}$, where $Z_{\odot}$ is the solar metallicity, and
$f'=0.5$. Note that only the product $Zf$ can be studied in our approach.
For brevity, we hereafter fix $f=0.5$ and constrain the metallicity
profile $Z(r)$ below. However, one should keep in mind that what is really
constrained is $Z(r)f(r)/0.5$.

A general parametrization of the density radial dependence, which we adopt
in our study as well, is the so-called ``beta profile'',
\begin{equation}
n_{\rm e} (r) =n_{0} \left(1+\left(r/r_{\rm c} \right)^2   \right)^{-3
\beta/2}.
\label{Eq:n_e}
\end{equation}
It has three parameters, the normalization $n_{0}$, the slope $\beta$ and
the inner cutoff radius $r_{c}$. We are primarily interested in the outer
parts of the corona (at least, outside the Galaxy), where the dependence
on $r_{c}$ is negligible, and the electron density reduces to $n_{\rm
e}(r) \simeq n_{0} r_{c}^{3\beta} r^{-3\beta}$, so that the normalization
parameter is now $n_{0} r_{c}^{3\beta}$. It is this parameter which is
normally constrained by observations, and it is used in what follows.
Whenever a particular value of $r_{c}$ is needed, $r_{c}=3$~kpc is used,
though a change in this parameter has a negligible impact on the results.

\begin{figure}
\begin{center}
\includegraphics[width=0.95 \columnwidth]{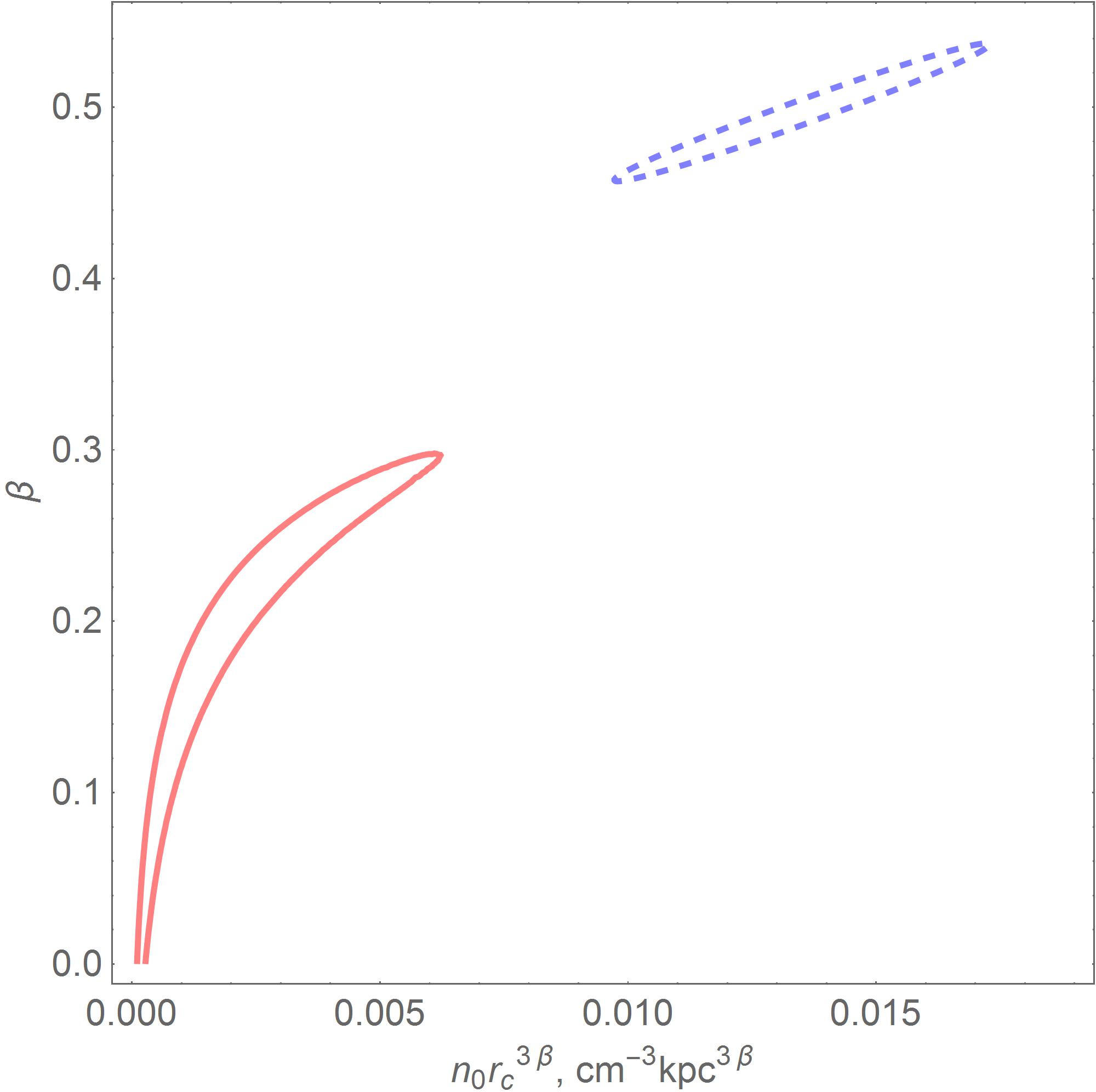}
\caption{
\label{fig:2contours}
Parameters of the density profile of the circumgalactic gas,
Eq.~(\ref{Eq:n_e}), 68\% CL contours. Red full line: combination of
ram-stripping constraints from nearby Milky-Way satellites (this work, see
text). Blue dashed line: combination of X-ray spectroscopic observations
assuming $Z=0.3 Z_{\odot}$ everywhere \citep{1412.3116}.}
\end{center}
\end{figure}

The reader may find the most recent discussion on constraining $n_{\rm
e}(r)$ from X-ray spectroscopy in \citet{1412.3116}, while more details of
the approach are presented in \citet{MB2013}. Assuming
$Z(r)f(r)=\mbox{const} =0.3 Z_{\odot} \times 0.5$, \citet{1412.3116}
obtain best-fit values of $\beta=0.50$ and $n_{0}
r_{c}^{3\beta}=0.0135$~cm$^{-3}$kpc$^{3\beta}$ (note a factor of 100
misprint in their Abstract). The corresponding 68\% allowed region in the
parameter space, see their Fig.~5, is shown as a dashed contour in
Fig.~\ref{fig:2contours}.

\subsection{Ram--pressure stripping}
\label{sec:ram}
Another approach is based on the observation  \citep{dwarfs250kpc} that
the amount of gas in dwarf satellites residing within $\sim 250$~kpc from
the centers of the Milky Way and M31 galaxies is much smaller than
that observed in more distant dwarfs. This observation is
interpreted in terms of the ram-pressure stripping due to travelling of
the dwarf galaxies through the circumgalactic gas of their giant
companions (see, however, \citep{1605.02746}, where other potential
contributions are discussed). Several constraints on the density of
the circumgalactic gas coming from observations of the Milky-Way satellites
are collected in Table~\ref{tab:dwarfs}.
\begin{table*}
\centering
\begin{minipage}{126mm}
\begin{tabular}{|ll|c|c|c|}
\hline
\hline
&Object & $r$, kpc& $n_{e}$, $10^{-4}$~cm$^{-3}$ & Reference\\
\hline
($+$)& Distant dwarfs& $\langle 0 \dots 250 \rangle$& $\ge 0.24$&
\citet{dwarfs250kpc}\\
&&(volume averaged)&&\\
\hline
&Carina dwarf& 20 (3 \dots 63)&0.85 (0.55 \dots 3.9)& \citet{0901.4975}\\
&Ursa Minor dwarf& 40 (10 \dots 76)& 2.1 (0.13 \dots 0.72)&\\
&Sculptor dwarf& 68 (31 \dots 83) & 2.7 (0.51 \dots 3.9)&\\
(*)& Fornax dwarf& 118 (66 \dots 144)& 3.1 (0.98 \dots 4.6)&\\
\hline
(*)& Sextans dwarf& 73.5 (59.8 \dots 90.2)& (1.3 \dots 5)&
\citet{1305.4176}\\
(*)& Carina dwarf& 64.7 (51.2 \dots 81.8)& (1.5 \dots 3.6)&\\
\hline
(*) & LMC& 48.2$\pm$5& 1.1$^{+0.44}_{-0.45}$&\citet{1507.07935}\\
\hline
($+$)& LMC pulsars& $\langle 0 \dots 50 \rangle$&$\le 5$&
\citet{DM}\\
 &(dispersion measure)&(line averaged)&&\\
\hline
\end{tabular}
\caption{\label{tab:dwarfs}
Direct constraints on the gas density (ram-pressure stripping and pulsar
dispersion measures). In the first column, (*) indicates that the
constraint was used in the statistical analysis performed in the present
work; $(+)$ indicates that this constraint is shown separately in
Fig.~\ref{fig:density-marg}. Round brackets denote uncertainty, angle
brackets denote averaging. }
\end{minipage}
\end{table*}
The dominant source of uncertainty here is the orbit of a satellite, which
determines the pericenter distance, where the stripping is most efficient.
To quantify the ensemble of these results, the standard
likelihood function, $\mathcal{ L}_{s}$, in terms of the two parameters of
the gas density distribution, $\beta $ and $n_0$, is constructed.
Observations with very large uncertainties almost do not affect the result
and are not taken into account, see Table~\ref{tab:dwarfs}. The resulting
68\% C.L.\ contour in the parameter space is shown with the full line in
Fig.~\ref{fig:2contours}. Note that this method constrains the gas density
directly, without invoking any assumption on the metallicity. The
constraint of \citet{dwarfs250kpc}, formulated in a different way (an
inequality based on a combined study of a sample of objects, bounding the
volume-averaged gas density from below), is not included in the fit. We
will return to this constraint in Sec.~\ref{sec:comb}.

\subsection{Other constraints}
\label{sec:DM}
For completeness, recall one more way to constrain the gas density
directly. It relies on the observation of pulsar dispersion measures which
are related to the column density of electrons along the line of sight.
\citet{DM} obtained an upper limit on the electron density integrated up
to the LMC, assuming a certain contribution from the inner part of the
Milky Way. We will return to this constraint in Sec.~\ref{sec:comb}.
Another approach, more robust with respect to assumptions about the
inner Galactic gas, was based on individual
pulsar studies \citep{Yana}.

\section{Combined constraints on the density and metallicity}
\label{sec:comb}
Figure~\ref{fig:2contours} looks disappointing at first sight since the
allowed parameter regions derived in two different approaches do not
overlap. However -- and this is the main point of the present Letter, --
relaxing the assumption of the constant metallicity $Z=0.3 Z_{\odot}$
brings them into agreement. Moreover, this opens a possibility to
constrain, for the first time, the metallicity profile of the
circumgalactic gas from observations.

To proceed further, we determine two likelihood functions corresponding to
the two approaches: one is the $\mathcal{ L}_{s}(n_{0},\beta )$ determined
in Sec.~\ref{sec:ram} and describing constraints from ram-pressure
stripping; another one, $\mathcal{ L}_{X}$, is responsible for the X-ray
constraints. The latter function should depend on parameters of the
metallicity profile besides those of the density profile. To quantify this
dependence, we assume a similar ``beta profile'' for the metallicity,
\begin{equation}
Z =A \left(1+\left(r/r_{\rm c} \right)^2   \right)^{-3
B/2}
\approx
A \left(r/r_{\rm c} \right)^{-3 B}.
\label{Eq:metall}
\end{equation}
At large distances from the Galactic Center, a straightforward comparison
of Eqs.~(\ref{Eq:*}), (\ref{Eq:n_e}) and (\ref{Eq:metall}) results in
simple relations between the true $(n_0, \beta )$ and reported (under
assumption of a certain metallicity; $n_{0}', \beta '$) parameters of the
electron densities,
$$
n_0 r_c^{3\beta }= \frac{1}{A} n_0' r_c^{3 (\beta '-B)},
$$
$$
\beta = \beta'-B.
$$
This allows one to generalize the likelihood $\mathcal{ L}'_{X} (n_0',
\beta ')$ to the more general
$\mathcal{ L}_{X} (n_{0}, \beta ,A,B)$. The original $\mathcal{ L}'_{X}$
function is assumed to be a deformed Gaussian reproducing the 68\% C.L.\
contour of \citet{1412.3116}, that is the dashed contour in
Fig.~\ref{fig:2contours}, correctly.

The standard statistical procedure allows one to determine combined
constraints on the parameter pairs $(n_{0},\beta )$ and $(A, B)$ by
marginalization (see e.g.\ \citet{marg}). Let $\mathcal{ L}_{e} (n_{0},
\beta )$ and $\mathcal{ L}_{Z} (A,B)$ be marginalized likelihood functions
for the parameters of gas density and metallicity, respectively. Then
$$
\mathcal{ L}_Z(A,B) = \int\!dn_0\,d\beta \, \mathcal{ L}_{X} (n_{0}, \beta
,A,B)   \mathcal{ L}_{s}(n_{0},\beta ),
$$
$$
\mathcal{ L}_e(n_0,\beta ) = \int\!dA\,d
B \, \mathcal{ L}_{X} (n_{0}, \beta
,A,B)   \mathcal{ L}_{s}(n_{0},\beta ).
$$
With the help of these new likelihood functions, best-fit values and 68\%
C.L.\ contours for combined constraints are easily determined.

\begin{figure}
\begin{center}
\includegraphics[width=0.95 \columnwidth]{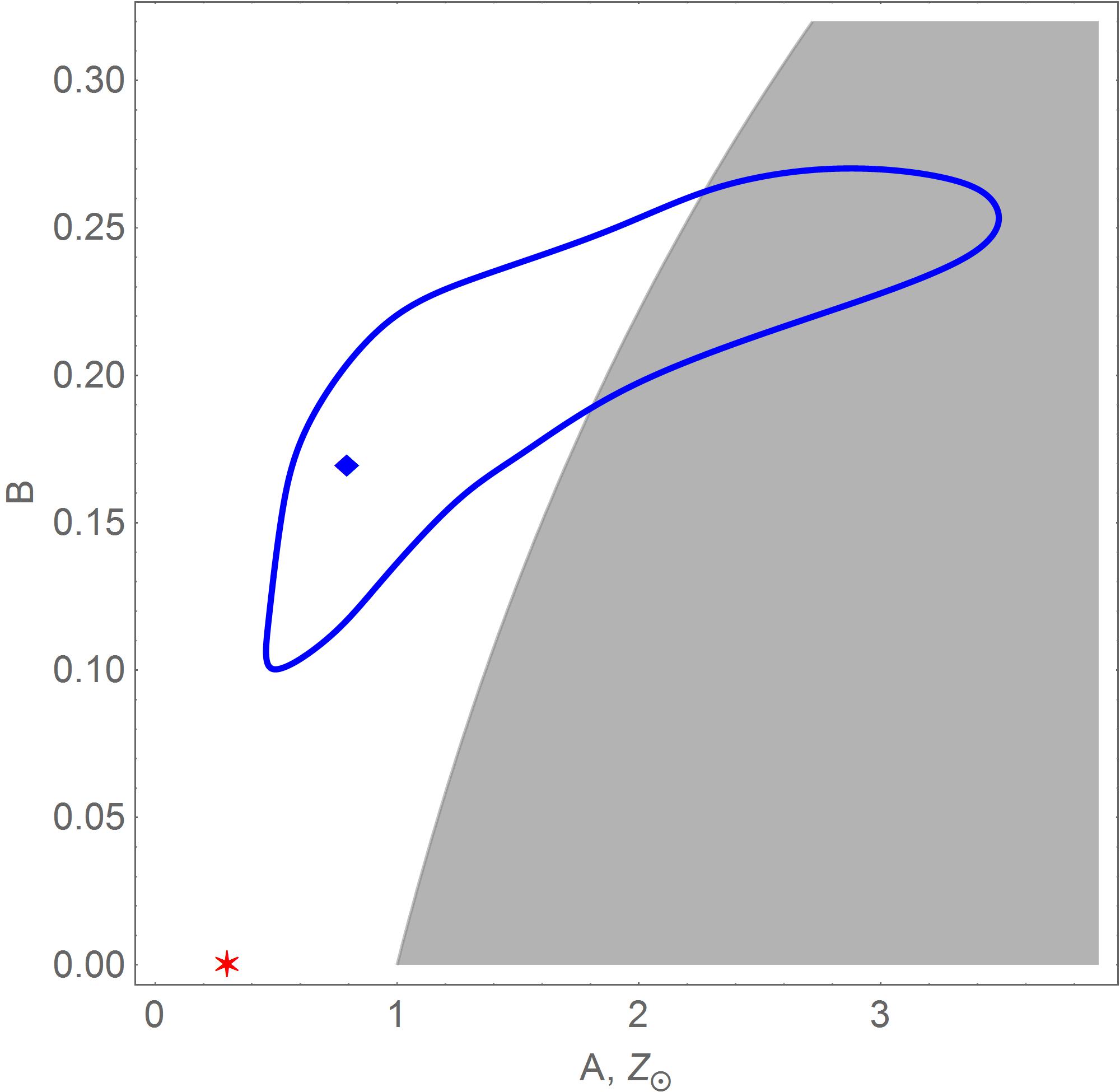}
\caption{
\label{fig:ABmarg}
Parameters of the metallicity profile of the circumgalactic gas,
Eq.~(\ref{Eq:metall}). The blue diamond and the blue contour represent,
respectively, the best-fit point and the 68\% CL contour obtained in this
work. The red asterisk gives the values assumed in previous studies. The
shaded area corresponds to profiles having $Z(r_{\odot}) \ge Z_{\odot}$
for the galactocentric distance of the Sun $r_{\odot}=8.5$~kpc. }
\end{center}
\end{figure}

\begin{figure}
\begin{center}
\includegraphics[width=0.95 \columnwidth]{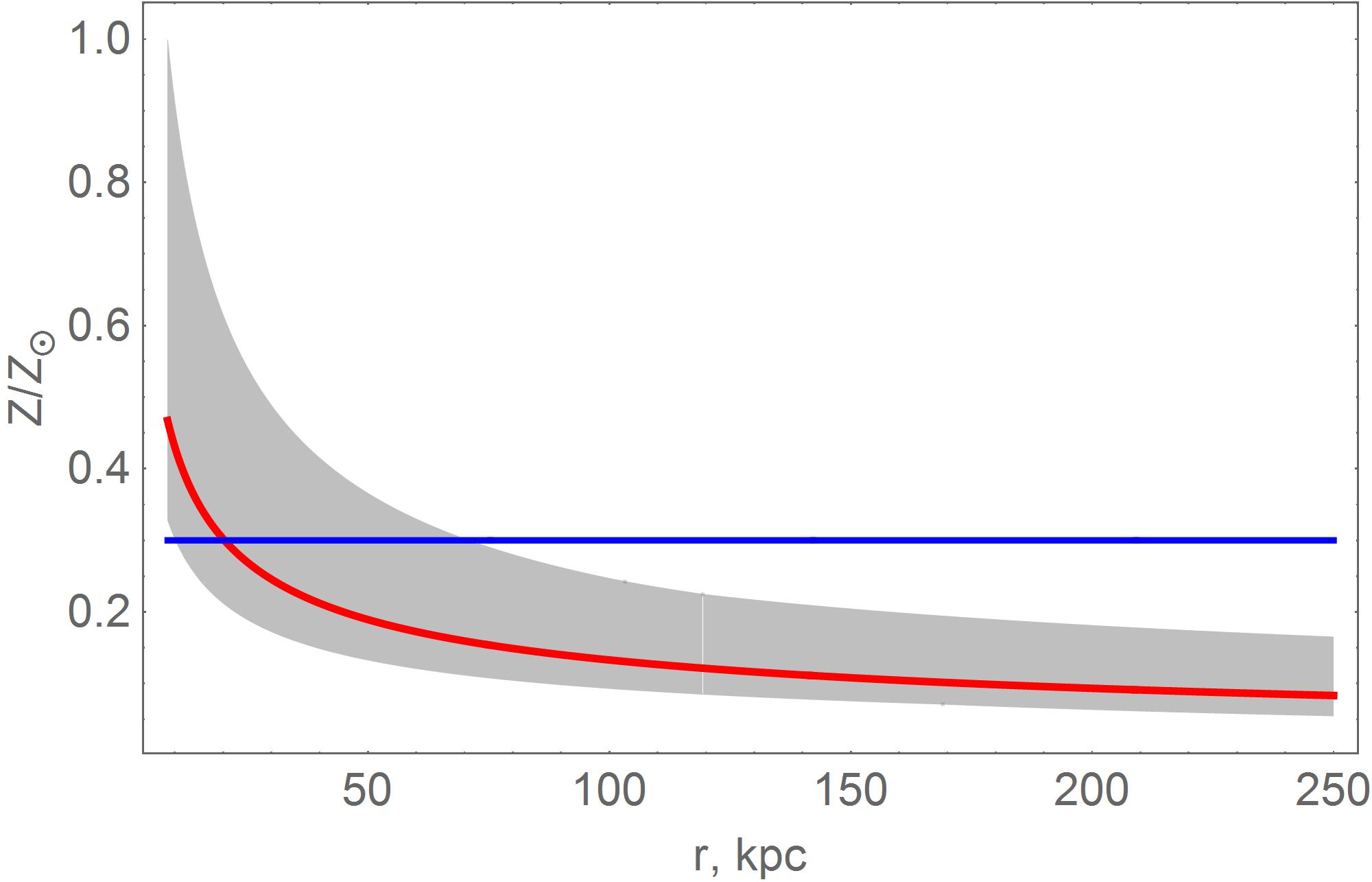}
\caption{
\label{fig:met-profiles}
Metallicity profiles of the circumgalactic gas. The red line and the gray
shaded area represent, respectively, the best-fit profile and 68\% CL
allowed range obtained in this work. The blue horizontal line is the
constant profile assumed in previous studies.}
\end{center}
\end{figure}

\begin{figure}
\begin{center}
\includegraphics[width=0.95 \columnwidth]{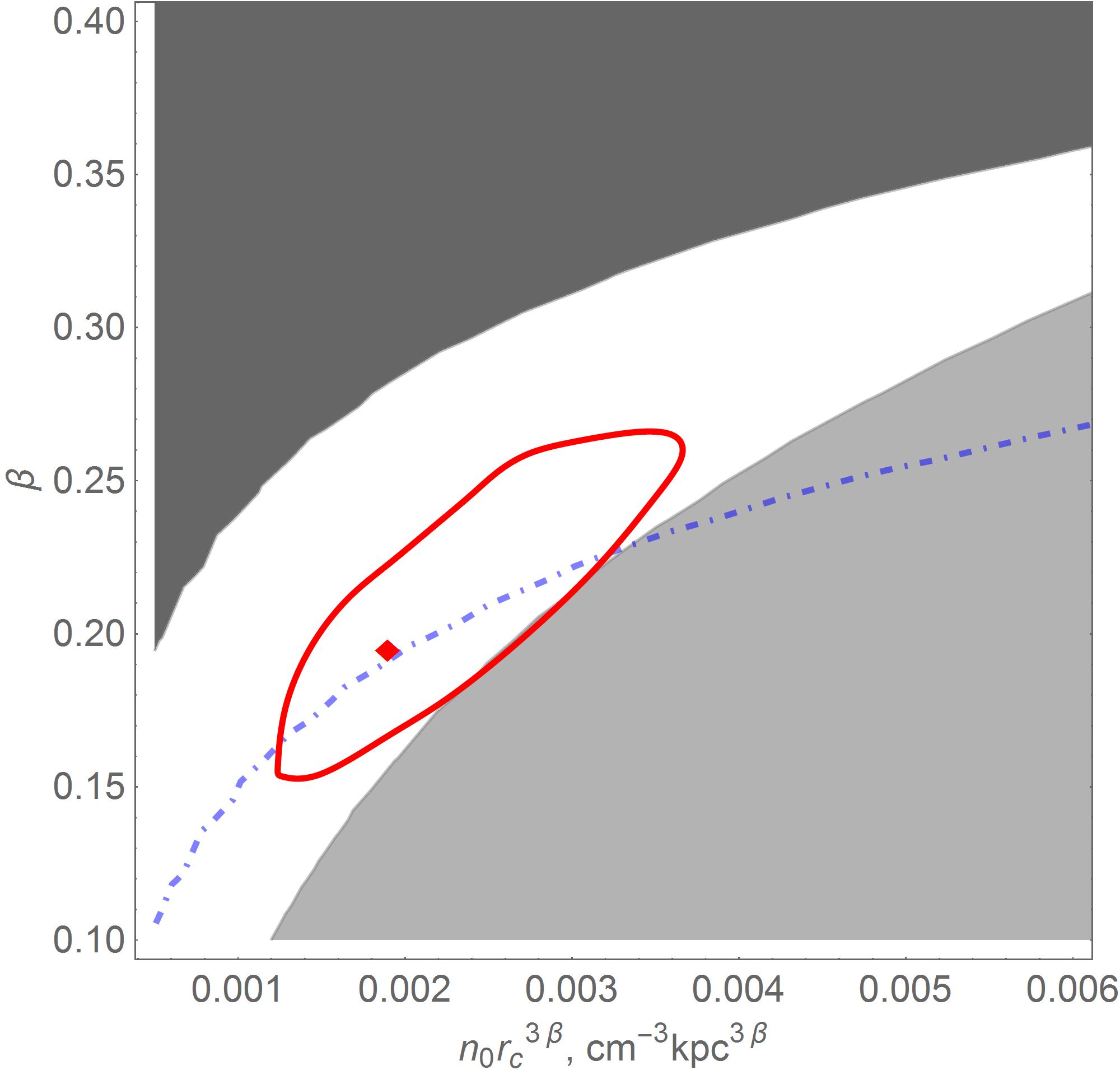}
\caption{
\label{fig:density-marg}
Constraints on the parameters of the density profile of the circumgalactic
gas, Eq.~(\ref{Eq:n_e}).
The red diamond and the red contour represent,
respectively, the best-fit point and the 68\% CL contour obtained in this
work.
The dark shaded area is excluded by \citet{dwarfs250kpc} from studies
of the ram-pressure stripping of distant dwarf satellites. The light
shaded area is excluded by \citet{DM} from the LMC pulsar dispersion
measures. The blue dash-dotted line corresponds to the total mass of the
gas of $1.6\times 10^{11} M_{\odot}$ required to explain $\sim 100\%$ of
the ``missing baryons''.
}
\end{center}
\end{figure}

\begin{figure}
\begin{center}
\includegraphics[width=0.95 \columnwidth]{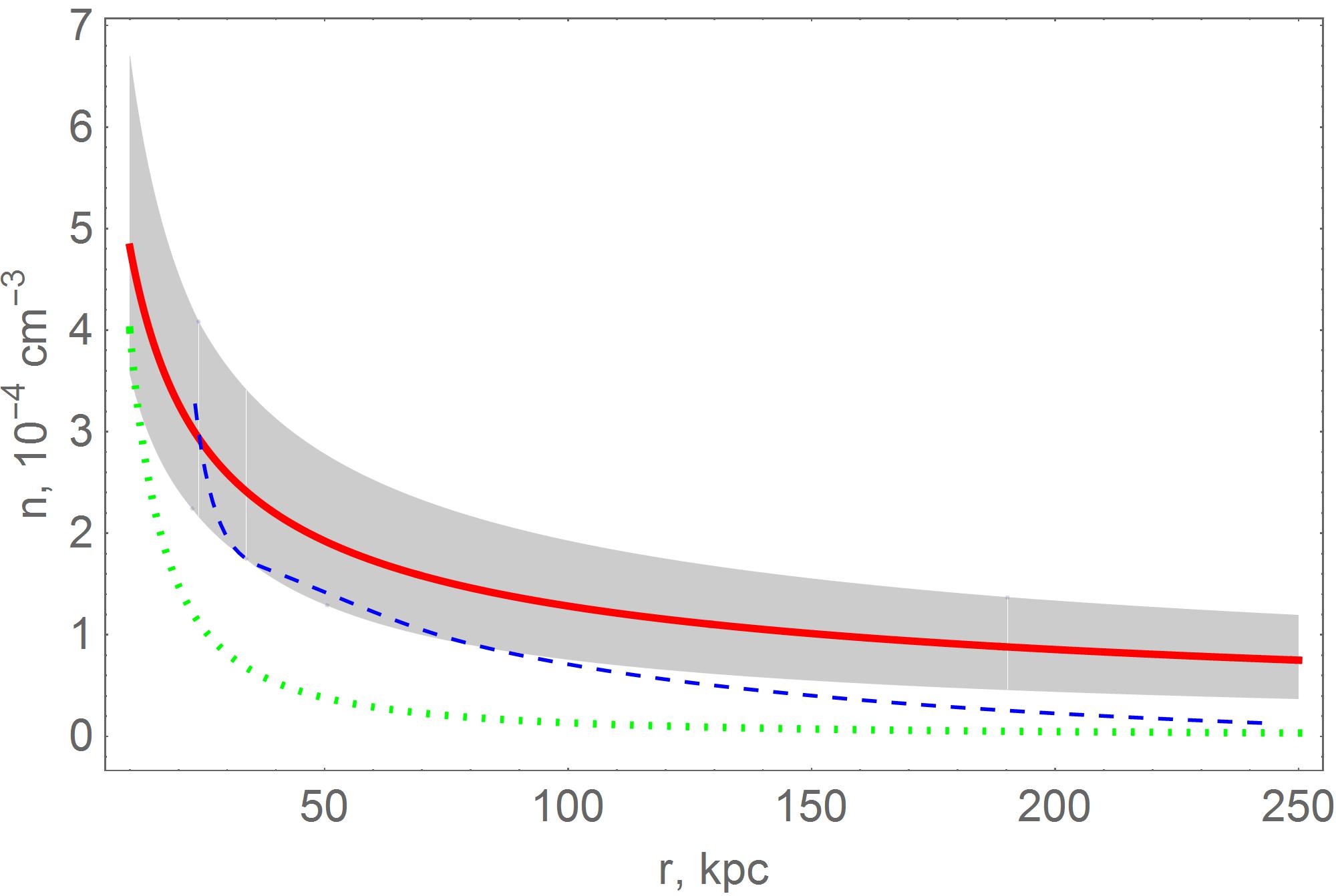}
\caption{
\label{fig:density-profiles}
Density profiles of the circumgalactic gas. The red line and the gray
shaded area represent, respectively, the best-fit profile and 68\% CL
allowed range obtained in this work.
The green dotted line is the best-fit profile of \citet{1412.3116}.
The blue dashed line is the
profile obtained by \citet{Hooper} by numerical simulations.
}
\end{center}
\end{figure}

Resulting constraints on the metallicity profile, Eq.~(\ref{Eq:metall}),
are shown in Fig.~\ref{fig:ABmarg}.
The shaded area there corresponds to the parameters resulting in the
metallicity at the solar location exceeding $Z_{\odot}$. This information
is indicative only since our study deals with the gas at much larger
distances from the Galactic Center. The corresponding range of metallicity
profiles is shown in Fig.~\ref{fig:met-profiles}.
Figure~\ref{fig:density-marg}
presents our constraints on the parameters of the density profile,
Eq.~(\ref{Eq:n_e}), together with the regions excluded by
\citet{dwarfs250kpc} and \citet{DM}. Our 68\% C.L.\ allowed region
satisfies their constraints. The corresponding range of density profiles
is shown in Fig.~\ref{fig:density-profiles}.

\section{Discussion and conclusions}
\label{sec:concl}
The results of this work eliminate the apparent discrepancy, see
Fig.~\ref{fig:2contours}, in parameters of the Milky-Way circumgalactic
gas estimated from X-ray spectroscopy and from studies of ram-pressure
stripping of Galactic dwarf satellites. A combined analysis of the
observational constraints, performed here, determines the range of the
allowed metallicity profiles of the gas residing up to 250~kpc from the
Galactic Center (Figs.~\ref{fig:ABmarg}, \ref{fig:met-profiles}). Not
surprisingly, the metallicity decreases considerably in the outer parts of
the halo \red with respect to the inner part. The profile in the outer
part is, however, fairly flat. \black

Importantly, we obtained constraints on the parameters of the gas density
distribution from a combination of all available data
(Figs.~\ref{fig:density-marg}, \ref{fig:density-profiles}). Compared to
the profile of \citet{1412.3116}, the best-fit one is flatter, resulting
in higher gas density in the peripheral parts of the Galactic corona and,
consequently, in larger total gas mass. This agrees well with recent
simulations \citep{HalfHidden} and ultraviolet O{\small VI} observations
\citep{1602.00689} indicating that the X-ray observations may underestimate
the total amount of circumgalactic gas by a factor of two. It is
interesting to compare the total gas mass with the ``missing baryon'' mass
of the Galaxy, since the halo of circumgalactic gas was suggested as an
explanation of the mismatch between the Milky-Way and cosmological average
baryon content \citep{DM}. One can see from Fig.~\ref{fig:density-marg},
where a line corresponding to the required ``missing-baryon'' mass is
shown, that our results support this explanation.

\begin{figure}
\begin{center}
\includegraphics[width=0.95 \columnwidth]{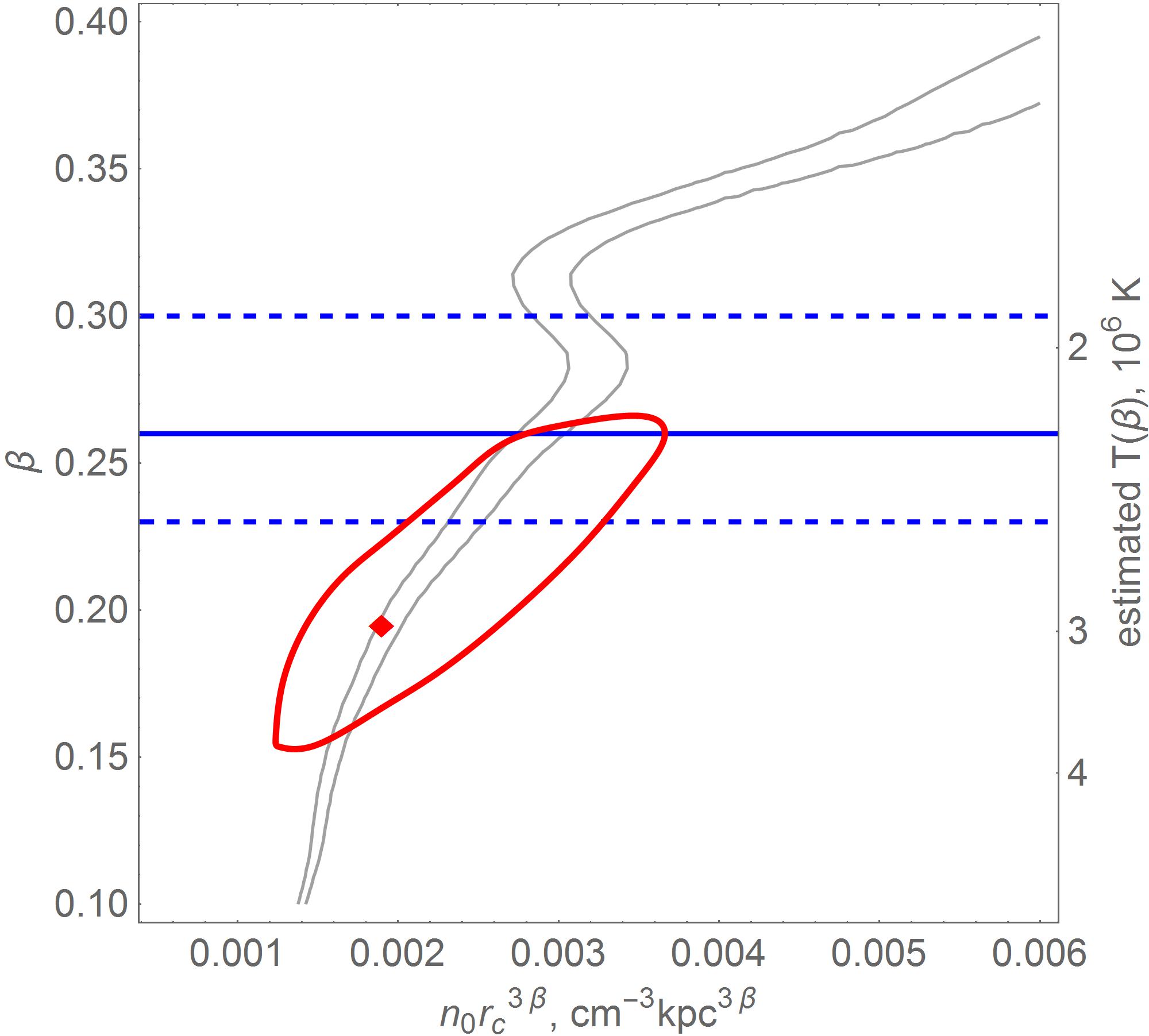}
\caption{
\label{fig:T}
\red
Estimates of the halo temperature $T$ and X-ray luminosity $L_{X}$ versus
the density profile parameters.
The red diamond and the red contour represent,
respectively, the best-fit point and the 68\% CL contour obtained in this
work.
The right scale represents $T$ estimated from $\beta$.
Blue horizonthal lines give
the $T$ median value (full line, $2.22 \times 10^{6}$~K) and
interquantile range (dashed lines, $0.63 \times 10^{6}$~K) from
\citet{1306.2312}.
Thin gray lines bound the range $L_{X}=(2\dots 3)\times 10^{39}$~erg/s
favoured by  \citet{ApJ_485_125,a-p/9710144}.
See the text for details and important notes. \black
}
\end{center}
\end{figure}

\red
It is also interesting to see how our density profile agrees with estimates
of the temperature and the luminosity of the Galactic X-ray halo. To this
end, we note the relation between the slope parameter $
\beta$ in
Eq.~(\ref{Eq:n_e}), the velocity dispersion of galactic objects $\sigma$
and the gas temperature $T$,
\begin{equation}
\beta=\frac{\mu m_p \sigma^2}{kT},
\label{Eq:T}
\end{equation}
see e.g.\ \citet{NFW}, where $\mu$ is the mean atomic mass per particle,
$m_{p}$ is the proton mass and $k$ is the Boltzmann constant. The velocity
dispersion is consistent with $\sigma \sim 100$~km/s in the inner $\sim
80$~kpc \citep{1304.5127}, but several observations indicate significant
decrease of $\sigma$ at large $r$ \citep{a-p/0506102,0910.2242}. We use
$\sigma=90$~km/s in the following estimates. The estimated values of $T$
are shown in the right scale of Fig.~\ref{fig:T}.
Given the approximate nature of these estimates, our density profiles are
in a good agreement with observational constraints on the X-ray
temperature of the halo gas, for instance, those by \citet{1306.2312},
shown in Fig.~\ref{fig:T}. We also calculate the total X-ray luminosity of
the halo, $L_{x}$, by making use of Eqns.~(17)-- (19) of
\citet{1412.3116}. The metallicity enters there through the cooling
function \citep{ApJ_88_253} which we take for [Fe/H]$=-1$, corresponding
to the most part of the halo, cf.\ Fig.~\ref{fig:met-profiles}.
Observations point to $L_{X}\sim (2\dots 3)\times 10^{39}$~erg/s
\citep{ApJ_485_125,a-p/9710144}, the range shown in Fig.~\ref{fig:T} as
well, again in a good agreement with our preferred parameters for the
density profile.

However, these \blue temperature-related \black estimates should be
considered with caution, because of several reasons. Firstly, the
constraints discussed in this paper are relevant for the outer part of the
halo, where observational information on X rays is scarce. Secondly, the
estimates assume that the halo is isothermal while some studies point to
the opposite \citep{0906.1532}. Thirdly, they assumed constant metallicity
and velocity dispersion. Finally, the quantitative values of temperature
and luminosity depend on the values of poorly known parameters to which
the results of the present paper are insensitive, like $r_{c}$ and
$\sigma$\blue, for which we have very little data to work with\black.
Therefore, these considerations and results presented in Fig.~\ref{fig:T}
should be considered only as a demonstration of the qualitative agreement
of our model with observational data on $T$ and $L_{X}$.
\blue Indeed, e.g., the replacement of the velocity-dispersion temperature
in Eq.~(\ref{Eq:T}) by the rotation-curve based temperature estimate would
change the gas temperature by a factor of $\sim 2$, indicating a factor of
$\sim 2$ higher $\beta$; however, the accuracy of Eq.~(\ref{Eq:T}) is of
the same order.

Clearly, the beta models themselves might be very crude tools for
modelling of the possibly structured circumgalactic gas medium, but
relaxing the constant-metallicity assumption and removal of the
discrepancies we discuss here are necessary first steps towards
understanding of this interesting part of the Galaxy. The gas density
profile obtained in this work is considerably flatter than the total mass
density profile of the halo, in qualitative agreement with simulations by
\black
\citet{Hooper}. At galactocentric distances
$\lesssim 40$~kpc, the profile of \citet{Hooper} deviates from the beta
profile, Eq.~(\ref{Eq:n_e}), towards higher densities. Unfortunately, this
range of distances is not controlled by our approach; therefore, higher
densities are not experimentally excluded there.

\section*{Acknowledgments}

The author is indebted to O.~Kalashev and M.~Pshirkov
for interesting discussions
and to G.~Rubtsov for many helpful remarks. This work was supported
by the Russian Science Foundation (grant 14-12-01340).

\bsp

\label{lastpage}


\begin{thebibliography}{23}
\bibitem[\protect\citeauthoryear{Anderson \& Bregman}{2010}]{DM}
Anderson M.\ E., Bregman J.\ N.,
2010, ApJ, 714, 320

\bibitem[\protect\citeauthoryear{Battaglia et.\ al.}{2005}]{a-p/0506102}
\red Battaglia G.\ {\it et al.},
2005, MNRAS, 364, 433;
Erratum:
2006, MNRAS, 370, 1055
\black
\bibitem[\protect\citeauthoryear{Blitz \& Robishaw}{2000}]{dwarfs250kpc}
Blitz L., Robishaw T.,
2000, ApJ, 541, 675

\bibitem[\protect\citeauthoryear{Brown et.\ al.}{2010}]{0910.2242}
\red Brown W.\ R., Geller M.\ J., Kenyon S.\ J., Diaferio A.,
2010, ApJ, 139, 59.
\black
\bibitem[\protect\citeauthoryear{Cowan}{2013}]{marg}
  Cowan G.,
  2013, arXiv:1307.2487 [hep-ex].

\bibitem[\protect\citeauthoryear{Emerick et al.}{2016}]{1605.02746}
Emerick A., Mac Low M.-M., Grcevich J., Gatto A.,
2016, ApJ, 826, 148

\bibitem[\protect\citeauthoryear{Faerman, Sternberg \&
McKee}{2016}]{1602.00689} Faerman Y., Sternberg A., McKee C.\ F.,
2016, arXiv:1602.00689.

\bibitem[\protect\citeauthoryear{Feldmann, Hooper \&
Gnedin}{2013}]{Hooper}
Feldmann R., Hooper D., Gnedin N.\ Y.,
2013, ApJ, 763, 21

\bibitem[\protect\citeauthoryear{Gatto et al.}{2013}]{1305.4176}
 Gatto A., Fraternali F., Read J.\ I., Marinacci F., Lux H., Walch S.,
2013, MNRAS, 433, 2749

\bibitem[\protect\citeauthoryear{Grcevich \& Putman}{2009}]{0901.4975}
Grcevich J., Putman M.\ E.,
2009, ApJ, 696, 385;
   Erratum:
2010, ApJ, 721, 922

\bibitem[\protect\citeauthoryear{Gupta et al.}{2012}]{gas1}
Gupta A., Mathur S., Krongold Y., Nicastro F., Galeazzi M.,
2012, ApJ, 756, L8

\bibitem[\protect\citeauthoryear{Henley \& Shelton}{2013}]{1306.2312}
 \red Henley D.\ B., Shelton R.\ L.,
2013, ApJ, 773, 92
\black
\bibitem[\protect\citeauthoryear{Lei, Shelton \& Henley}{2009}]{0906.1532}
 \red Lei S., Shelton R.\ L., Henley D.\ B.,
2009, ApJ, 699, 1891
  \black
\bibitem[\protect\citeauthoryear{Miller \& Bregman}{2013}]{MB2013}
Miller M.\ J., Bregman J.\ N.,
2013, ApJ, 770, 118

\bibitem[\protect\citeauthoryear{Miller \& Bregman}{2015}]{1412.3116}
Miller M.\ J., Bregman J.\ N.,
2015, ApJ, 800, 14

\bibitem[\protect\citeauthoryear{Navarro, Frenk \& White}{1996}]{NFW}
\red Navarro J.\ F., Frenk C.\ S., White S.\ D.\ M.,
1996, ApJ, 462, 563
\black
\bibitem[\protect\citeauthoryear{Nesti \& Salucii}{2013}]{1304.5127}
 \red Nesti F., Salucci P.,
2013,  JCAP, 1307, 016
\black
\bibitem[\protect\citeauthoryear{Nugaev, Rubtsov \& Zhezher}{2016}]{Yana}
Nugaev E.\ Ya., Rubtsov G.\ I., Zhezher Ya.\ V.,
2016, Astron.\ Lett., 42, 173

\bibitem[\protect\citeauthoryear{Salem et al.}{2015}]{1507.07935}
Salem M., Besla G., Bryan G., Putman M., van der Marel R.\ P., Tonnesen,
S.,
2015, ApJ, 815, 77

\bibitem[\protect\citeauthoryear{Snowden et al.}{1997}]{ApJ_485_125}
\red Snowden S.\ L.\ {\it et al.},
1997, ApJ, 485, 125.
\black
\bibitem[\protect\citeauthoryear{Sutherland \& Dopita}{1993}]{ApJ_88_253}
\red Sutherland R.\ S., Dopita M.\ A.,
1993, ApJS, 88, 253.
\black
\bibitem[\protect\citeauthoryear{Taylor, Gabici \& Aharonian}{2014}]{Aha}
Taylor A.\ M., Gabici S.\ and Aharonian F.,
2014, PRD, 89, 103003

\bibitem[\protect\citeauthoryear{Wang}{1998}]{a-p/9710144}
\red Wang Q.\ D.,
1998, Lect.\ Notes Phys.,  506, 503
\black
\bibitem[\protect\citeauthoryear{Zheng et al.}{2015}]{HalfHidden}
Zheng Y., Putman M.\ E., Peek J.\ E.\ G.. Joung M.\ R.,
2015, ApJ, 807, 103
\end{thebibliography}
\end{document}